\documentclass[a4paper,12pt]{article}
\usepackage{geometry}                
\usepackage{graphicx}
\usepackage{amssymb}
\usepackage{epstopdf}
\usepackage{color}

\newcommand{\us}{{\underline{s}}}

\title{On sampling and modeling complex systems}
\author{Matteo Marsili%
  \thanks{The Abdus Salam International Centre for Theoretical
    Physics, Strada Costiera 11, 34014 Trieste, Italy}, \
    Iacopo Mastromatteo%
  \thanks{Capital Fund Management, 21-23 Rue de l'Universit\'e, 75007 Paris, France}
 \ and Yasser Roudi%
  \thanks{Kavli Institute for Systems Neuroscience, NTNU, Trondheim, Norway
Nordita, KTH Royal Institute of Technology and Stockholm University, Stockholm, Sweden }}
\date{}                                           

\begin{document}
\maketitle
\begin{abstract}
The study of complex systems is limited by the fact that only few variables are accessible for modeling and sampling, which are not necessarily the most relevant ones to explain the systems behavior. In addition, empirical data typically under sample the space of possible states. 
We study a generic framework where a complex system is seen as a system of many interacting degrees of freedom, which are known only in part, that optimize a given function. We show that the underlying distribution with respect to the known variables has the Boltzmann form, with a temperature that depends on the number of unknown variables. In particular, when 
the influence of the unknown degrees of freedom on the known variables is not too irregular, 
the temperature decreases as the number of variables increases. This suggests that models can be predictable only when the number of relevant variables is {\em less} than a critical threshold. 
Concerning sampling, we argue that the information that a sample contains on the behavior of the system 
is quantified by the entropy of the frequency with which different states occur. This allows us to characterize the properties of {\em maximally informative samples}:
within a simple approximation, the most informative frequency size distributions have power law behavior and Zipf's law emerges at the crossover between the under sampled regime and the regime where the sample contains enough statistics to make inference on the behavior of the system. These ideas are illustrated in some applications, showing that they can be used to identify relevant variables or to select most informative representations of data, e.g. in data clustering.
\end{abstract}

\section{Introduction}
Complex systems such as cells, the brain, the choice behavior of an individual or the economy can generally be regarded as systems of many interacting variables. Their distinguishing feature is that, contrary to generic random systems, they perform a specific function and exhibit non-trivial behaviors. Quantitative science deals with collecting experimental or empirical data that reveal the inherent mechanisms and organizing principles that suffice to reproduce the observed behavior within theoretical models. The construction of machines or the design of intervention which achieve a desired outcome, as e.g. in drug design~\cite{drugdesign} or for the regulation of financial markets~\cite{Haldane}, crucially depend on the accuracy of the models.

This endeavor has intrinsic limits: our representations of complex systems are not only approximate, they are incomplete. They take into account only few variables -- that are at best the most relevant ones -- and the interactions among these. By necessity they neglect a host of other variables, that also affect the behavior of the system, even though on a weaker scale. These are not only variables we neglect, but {\em unknown unknowns} we do not even know they exist and have an effect. 

This is not necessarily a problem as long as {\em i)} the phenomenon depends on few relevant variables and {\em ii)} one is able to identify and to probe them. Indeed, as E. P. Wigner argues ``It is the skill and ingenuity of the experimenter which show him phenomena which depend on a relatively narrow set of relatively easily realizable and reproducible conditions"~\cite{Wigner}. 
Yet, even if advances in IT and experimental techniques have boosted our ability to probe complex systems to an unprecedented level of detail, we are typically in the situation where the state space of the system at hand is severely under sampled and relevant variables (e.g. the expression of a gene) are in many cases inferred from indirect measurements.

In addition, there are intriguing statistical regularities that arise frequently when probing complex systems. Frequency counts in large samples often exhibit the so-called Zipf's law,  according to which the $k^{\rm th}$ most frequent observation occurs with a frequency that is roughly proportional to $1/k$, an observation that has attracted considerable interest over several decades now\footnote{The literature on this finding is so vast that a proper account would require a treatise of its own. We refer to recent reviews~\cite{review_zipf} and papers~\cite{MoraBialek,Baek,pnerozipf} and references therein.}. 
Model systems in physics, e.g. for ferromagnetism, exhibit similar scale free behavior only at  special ``critical" points, where the system undergoes a phase transition. This leads to wonder about mechanisms by which nature would self-organize to a critical point~\cite{Bak} or on the generic features of systems that share this property~\cite{MoraBialek}. 
Yet, the fact that Zipf's law occurs in a wide variety of different systems, 
suggests that it does not convey specific information about the mechanism of self-organization of any of them. 


Here we address the general problem of modeling and sampling a complex system from a theoretical point of view. We focus on a class of complex systems which are assumed to maximize an objective function depending on a large number of variables. Only some of the variables are known, whereas the others are unknown. Accordingly, only the part of the function that depends solely on the known variables is known, for the rest one can at best know its statistics.  The assumption that complex systems optimize some function, even if it is widely used in modeling (e.g. utility/fitness maximization in economics/biology),  may be debatable. Still, it allows us to address two related issues: First, under what conditions do models based on a subset of known variables reproduce systems behavior? How many variables should our models account for and how relevant should they be? Second, can we quantify how much information a given sample contains on the behavior of a complex system? What is the maximal amount of information that a finite data set can contain and what are the properties of optimally informative samples in the strongly under sampled regime?

In section 2, after constructing a mathematically well defined set up, we first discuss the issue of model's predictability: given some knowledge about how the objective function depends on the observed variables, what is the probability that we correctly predict the behavior of these variables? We show that, under very broad conditions, the dependence of the probability to observe a given outcome 
on the (observable part of the) objective function takes a Gibbs-Boltzmann form. 
In particular, if the dependence on unknown variables is not too irregular -- i.e. if the distribution of the unknown part of the objective function has thin tails -- then the ``temperature'' parameter {\em decreases} with the number of unknown variables.
This suggests that, models are predictable only when the number of unknown variables is {\em large} enough. This is illustrated for a particular case, drawing from results on the Random Energy Model~\cite{REM}, which is worked out in the Appendix. There we find that models are predictable only when the number of known variables is {\em less} than a critical threshold. This suggests a general argument for the non-trivial fact that ``in spite of the baffling complexity of the world, [...] phenomena which are independent  of all but a manageably small set of conditions" exist at all, which makes science possible~\cite{Wigner}. 

In section 3 we will then be concerned with what can be called an inverse problem:  if we choose some variables to observe, and collect a number of samples, how much do we learn about the objective function? We argue that {\em i}) the information that the sample contains on the behavior of the system is quantified by the entropy of the frequency with which different states occur. On the basis of this, {\em ii}) we characterize most informative samples and we find that their frequency size distributions, in the under sampled regime, have power law behavior. Within our approximated treatment, we find that the under sampling regime can be distinguished from the regime where the 
sample 
contains enough statistics to make inference on the underlying distribution. Finally, {\em ii}) the distribution with the highest information content coincides with Zipf's law, which attains at the crossover between these two regimes.

Finally section 4 gives evidences, based on concrete applications in proteins, finance and language, that these insights can be turned into practical criteria for studying complex systems, in particular for selecting relevant variables and/or the most informative representation of them.


\section{The setup}
\label{sec:setup}

\begin{figure} 
\begin{center} 
\includegraphics[width=7cm]{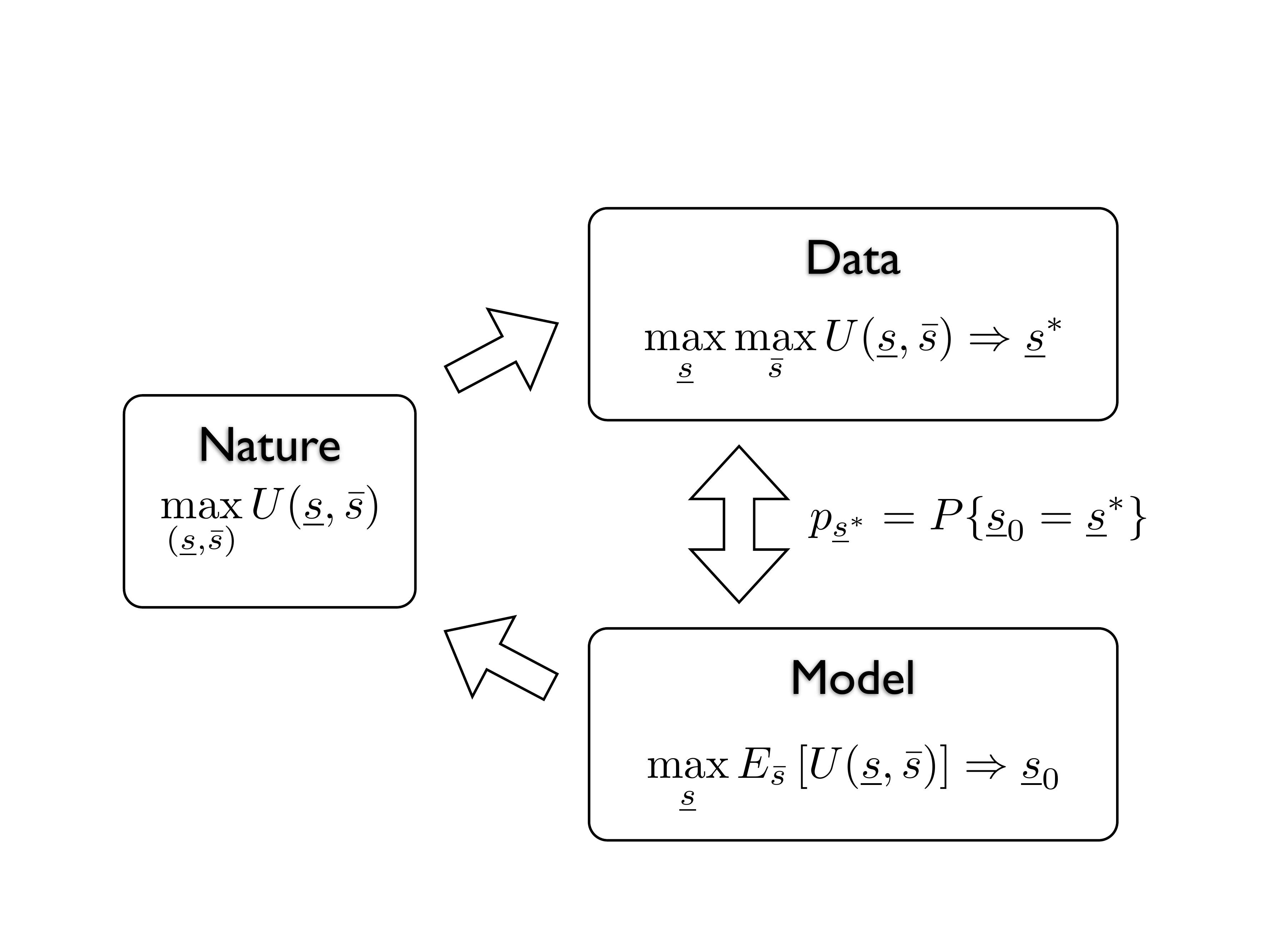} 
\caption{\label{fig1} Sketch of the setup: $\underline{s}$ are the known variables. The behavior of the system is encoded in the optimal choice $\underline{s}^*$. This results from the maximization of a function $U(\underline{s},\bar s)$ which also depends on unknown variables $\bar s$. Assuming it is possible to model the dependence of the objective function on the known variables $\underline{s}$, i.e. that $u_{\underline{s}}=E_{\bar s}[U(\underline{s},\bar s)]$ is known, what is the probability that the model's prediction $\underline{s}_0$ matches the observed behavior of the system? How relevant and how many should the known variable be?} 
\end{center} 
\end{figure} 

We consider a system which optimizes a given function $U(\vec s)$ over a certain number of variables $\vec s=(\underline s,\bar s)$. Only a fraction of the variables -- the ``knowns'' $\underline{s}$ -- are known to the modeler, as well as that part of the objective function $u_{\underline{s}}$ that depends solely on them. The objective function also depends on other variables $\bar s$ -- the ``unknowns'' -- in ways that are unknown to the modeler. Formally, we can define $u_{\underline{s}}=E_{\bar s}[U(\vec s)]$, where $E_{\bar s}[\ldots]$ stands for the expected value over a prior distribution on the dependence of $U(\vec s)$ on the unknown variables, that encodes our ignorance on them. In other words, 

\begin{equation}
\label{ }
U(\vec s)=u_{\underline{s}}+v_{\bar s|\underline{s}}
\end{equation}
where $v_{\bar s|\us}=U(\vec s)-E_{\bar s}[U(\vec s)]$ is an unknown function of $\bar s$ and $\us$, that we 
assume to be drawn randomly and independently for each $\vec s=(\us,\bar s)$ from a given distribution $p(v)$. Hence  
 $E_{\bar s}[\ldots]$ denotes the expectation with respect to this distribution.
The fact that $v_{\bar s|\us}$ are independent draws from $p(v)$ here translates in the fact that knowledge of $\bar s$ does not provide any information on $\us$ as long as $v_{\bar s|\us}$ is unknown. This is what would be dictated by the maximum entropy principle~\cite{Jaynes}, in the absence of other information on the specific dependence of $U$ on $\vec s$\footnote{Indeed, if the variables were not independent, we should have some information on their mutual dependence and if they were not identical we should have some clue of how they differ.}. In an information theoretic sense, this also corresponds to the most complex model we could think of for the unknown part of the system, as its full specification requires a number of parameters that grows exponentially 
with the number of unknown variables. 

Therefore, the behavior of the system is given by the solution 
\begin{equation}
\label{ }
\vec s^*=(\underline{s}^*,\bar s^*)\equiv{\rm arg}\max_{\vec s} U(\vec s)
\end{equation}
whereas the behavior predicted by the model, on the known variables, is given by 
\begin{equation}
\label{ }
\us_0\equiv{\rm arg}\max_{\underline{s}}u_{\underline{s}}. 
\end{equation}
Within this simplified description, the predictability of the model is quantified by the probability 
\begin{equation}
\label{ps}
p_{\underline s_0}=P\{\underline{s}_0= \underline{s}^*\} \equiv E_{\bar s}[\delta_{\us_0,\us^*}]
\end{equation}
that the model reproduces the behavior of the system. This setup is sketched in Fig.~\ref{fig1}. \\

Let us give few examples:
\begin{itemize}
  \item The choice of the city (i.e. $\us$) in which individuals decide to live, does not only depend on the characteristics of the city -- that may be encoded in some index $u_{\us}$ of city's living standards -- but also on unobserved factors ($\bar s$) in unknown individual specific ways. Here $v_{\bar s |\us}$ is a different function for each individual -- encoding the value of other things $\bar s$ he/she cares about (e.g. job and leisure opportunities, personal relations, etc), in the particular city $\us$.
  \item A plant selects its reproductive strategy depending on the environment where it leaves. This ends up in measurable phenotypic characteristics e.g. of its flowers,  that can be classified according to a discrete variables $\us$. The variables the species is optimizing over $\vec s=(\us,\bar s)$, 
also include unobserved variables $\bar s$, that influence other traits of the phenotype in unknown ways. 
  \item A text is made of words $\us$ in a given language. Each word $\us$ in the text has been chosen by the writer, depending on  the words $\bar s$ that precede and follow it, in order to efficiently represent  concepts in the most appropriate manner. We assume that this can be modeled by the writer maximizing some function $U(\vec s)$. 
  \item Proteins are not random hetero-polymers. They are optimized for performing a specific function, e.g. transmit a signal across the cellular membrane. This information is encoded in the sequence $\vec s$ of amino acids, however only a part of the chain ($\us$) is directly involved in the function (e.g. binding of some molecules at a specific site). The rest ($\bar s$) may have evolved to cope with issues that have nothing to do with the function, and that depend on the specific cellular environment the protein acts  in.
\end{itemize}



%

Within this set up, in the next section we address the following question: if we only have access to $\us$, how well can we predict the behavior of the system? More precisely what is the functional dependence of the probability for a configuration $\us$ to be the true maximum $\underline s^\star$?

\subsection{Gibbs distribution on $\us$}
\label{sec:Gibbs}

The functional dependence of the probability for a generic configuration $\us$ to be the true maximum $\underline s^\star$, which we have denoted as $p_{\us}=P\{\us=\us^* \}$, can be derived under very general conditions. 
We focus here on the case where all the moments are finite: $E_{\bar s}[v_{\bar s|\us}^m]<+\infty$ for all $m>0$. Without loss of generality, we can take $\underline{s}=(s_1,\ldots, s_n)$ and $\bar s=(s_{n+1},\ldots, s_N)$, with the variables $s_i=\pm 1$ taking two values for $i=1,\ldots,N$.  
The system would not be that complex if $n$ and $N$ were small, so we focus on the limit where both $n$ and $N$ are very large (ideally $n,N\to \infty$). 

For all ${\underline{s}}$, extreme value theory~\cite{Gumbel} shows that
\begin{equation}
\label{maxv}
\max_{\bar s} v_{\bar s|{\underline{s}}} \cong a+\frac{\eta_{\underline{s}}}{\beta},
\end{equation}
where $a$ is a constant, $\eta_{\underline{s}}$ are i.i.d. Gumbel distributed, i.e. $P\{\eta_{\underline{s}}<x\}=e^{-e^{-x}}$ and $\beta$ depends on the tail behavior of the distribution of $v_{\bar s|\us}$ (see later). Therefore
\begin{eqnarray}
p_{\underline{s}}&\equiv &P\{{\underline{s}} = {\underline{s}}^*\}=P\{\beta u_{\underline{s}}+\eta_{\underline{s}}\ge \beta u_{\underline{s}'}+\eta_{\underline{s}'},\forall {\underline{s}'}\neq {\underline{s}}\}\\
&=& \int_{-\infty}^\infty\!d\eta_{\underline{s}}e^{-\eta_{\underline{s}}-e^{-\eta_{\underline{s}}}}
\prod_{{\underline{s}'}\neq{\underline{s}}}\int_{-\infty}^{\eta_{\underline{s}}+\beta(u_{\underline{s}}-u_{\underline{s}'})}\!d\eta_{\underline{s}'}e^{-\eta_{\underline{s}'}-e^{-\eta_{\underline{s}'}}}\\
& = & \frac{1}{Z(\beta)}e^{\beta u_{\underline{s}}},\qquad Z(\beta)=
{\sum_{\underline{s}'}e^{\beta u_{\underline{s}'}}}\label{Logit}
\end{eqnarray}
which is the Boltzmann distribution, also called Logit model in choice theory. The derivation of the Logit model from a random utility model under the assumption of Gumbel distributed utilities is well known~\cite{McFadden,JPB}. Limit theorems on extremes dictate the form of this distribution for the whole class of models for which $v_{\bar s|{\underline{s}}}$ have all finite moments. 
This result extend to the case where $v_{\bar s|{\underline{s}}}$ are weakly dependent, as discussed in~\cite{Gumbel}.

The result of Eq.~(\ref{Logit}) could have been reached on the basis of maximum entropy arguments alone: 
On the true maximum, $\underline{s}^*$, the model's utility attains a value $u_{\underline{s}^*}$ that will generally be smaller than $u_{\underline{s}_0}$. Without further knowledge, the best prediction for $p_{\underline{s}}$ is given by the distribution of maximal entropy consistent with $E[u_{\underline{s}}]=u_{\underline{s}^*}$. 
It is well known that the solution of this problem yields a distribution of the form (\ref{Logit}). While this is reassuring, maximum entropy alone does not predict how the value of $\beta$ depends on the number of unknown unknowns. 
By contrast, extreme value theory implies that if the asymptotic behavior of $p(v)$ for large $v$ is given by $\log p(v) \sim -|v|^{\gamma}$, then one can take
\begin{equation}
\label{ }
\beta= \left[(N-n)\log 2\right]^{1-1/\gamma}
\end{equation}

One may na\"ively expect that the predictability of the model $p_{\us_0}$ gets smaller when the number $N-n$ of unknown variables increases. This is only true for $\gamma<1$, as indeed $\beta$ decreases as the number of unknown unknowns increases in this case.
When $p(v)$ decays faster than exponential ($\gamma>1$), which includes the case of Gaussian variables, $\beta$ diverges with the number of unknowns. If the number $n$ of observed variables stays finite, we expect that $p_{\us_0}\to 1$ in the limit of an infinite number of unknown variables. 

A manifestation of this non-trivial behavior is illustrated by the Gaussian case ($\gamma=2$) where also $u_{\us}$ are assumed to be i.i.d. draws from a Gaussian distribution with variance\footnote{$\sigma$ quantifies the relevance of the known variables. Note indeed that the typical variation $\Delta U$ of the objective function when a known variable is flipped is $\sqrt{1+\sigma^2}$ times larger than the change $\Delta U$ due to flipping an unknown variable. Hence known variables are also the most relevant ones.} $\sigma^2$. 
There, as shown in the appendix, for a given value of $\sigma$, the prediction of the model is reliable only as long as the fraction $f=n/N$ of known variables is {\em smaller} than a critical value $f_c=\sigma^2/(1+\sigma^2)$.

Summarizing, in this section we have shown that, given the form of $u_\us$, the probability to observe a certain state $\us$ follows a Gibbs-Boltzmann form with a ``temperature'' that depends on the number of unknown variables. 
A natural question one may ask at this point is the inverse problem to this: how much can we tell about $u_\us$ by observing the system? This is the question that we will address in the next section.

\section{Learning from sampling a complex system}

Given a sample $(\us^{(1)},\ldots,\us^{(M)})$ of $M$ observations of the state of a system, what can we learn on its behavior? As before, our working hypothesis is that $\us^{(i)}$ is the outcome of an optimization of an unknown function $U(\vec s)$ on a set of variables $\vec s$ that we observe only in part.  In order to connect to the {\em direct problem} discussed in the previous section, we note that one can also define $u_{\us}$, as $u_{\us}=E_{\bar s}[U(\vec s)]$, where the expected value, now, is an average over experiments carried out under the same experimental conditions, as far as the variables $\us$ are concerned. Therefore,  the function $u_{\us}$, while unknown, is the same across the sample.  The part of the objective function that depends on the unknown variables can again be defined as $v_{\bar s|\us}=U(\vec s)-u_{\us}$. However, since by definition there is no way to control for unknown variables, we cannot assume, {\em a priori}, that the influence of unknowns on the observed variables is the same across the sample. Rather, this is consistent with the function $v_{\bar s|\us}$ being a different independent draw from some distribution $p(v)$, for each $\vec s$ and for each point of the sample\footnote{Therefore we shall think of the sample as being the solution of the maximization problem: $\us^{(i)}={\rm arg}\max_{\us}\left[ u_{\us}+\max_{\bar s} v^{(i)}_{\bar s|\us}\right]$ for $i=1,\ldots,M$. For example, the choice of the city where Mr $i$ decides to live, also depends on individual circumstances, captured by the function $v^{(i)}_{\bar s|\us}$. Note furthermore that the number of unknown variables is assumed to be the same for all points of the sample. This implies that the unknown parameter $\beta$ in Eq.~(\ref{maxv}) is the same for all $i=1,\ldots,,M$.}.  Thus we shall think of the sample $(\us^{(1)},\ldots,\us^{(M)})$ as being $M$ independent configurations drawn from a distribution of the Gibbs-Boltzmann form as in Eq.~(\ref{Logit}). 

Let $K_{\us}$ be the number of times $\us$ was observed in the sample, that is

\begin{equation}
\label{ }
K_{\us} =\sum_{i=1}^M\delta_{\us^{(i)},\us} \; .
\end{equation}
In view of the discussion of the previous section, the relation between the 
distribution $p_{\us}$ that our data is sampling and the function $u_{\us}$ is given by the 
Gibbs-Boltzmann form of Eq.~(\ref{Logit}). This has two consequences:

\begin{enumerate}
  \item  Since the observed frequency $K_{\us}/M$ samples the unknown distribution $p_{\us}\sim e^{\beta u_{\us}}$, it also provides a noisy estimate of the unknown function
\begin{equation}
\label{ }
u_{\us}\approx c+\frac{1}{\beta}\log K_{\us}
\end{equation}
for some $c$ and $\beta>0$.
  \item Even without knowing what $u_{\us}$ is, we know that $p_{\us}$ is the maximal entropy distribution subject to an unknown constraint $E_{\us}[u]=\bar u$, or the distribution of maximal $E_{\us}[u]=\sum_{\us}p_{\us}u_{\us}$ with a given information content $H[\us]=\bar H$.
\end{enumerate}


The first observation highlights the fact that the information that we can extract from the sample on the function the system performs is given by the information contained in $K_{\us}$ and {\em not} in $\us$ itself. In order to make this observation more precise in information theoretic terms, we remark that, {\em a priori} all of the $M$ points $i$ in the sample should be assigned the same probability $P\{i\}=1/M$. With respect to this measure, the random variables $\us$ and $K_{\us}$ acquire  
distributions, respectively, given by $P\{\us^{(i)}=\us\}=K_{\us}/M$ and 
$P\{K_{\us^{(i)}}=k\}=k m_k/M$ where 
\begin{equation}
\label{ }
m_k=\sum_{\us}\delta_{k,K_{\us}}
\end{equation}
is the number of states $\us$ that are sampled exactly $k$ times.
Therefore their associated entropies are:
\begin{eqnarray}
\hat H[\us] & = & -\sum_{\us}\frac{K_{\us}}{M}\log \frac{K_{\us}}{M} =-\sum_k \frac{km_k}{M}\log \frac{k}{M}\\
\hat H[K] & = & - \sum_k\frac{km_k}{M}\log\frac{km_k}{M}=\hat H[\us]- \sum_k\frac{km_k}{M}\log m_k
\end{eqnarray}
where the notation $\hat H$ denotes empirical entropies. Since $K_{\us}$ is a noisy observation of the function $u_{\us}$, we conclude that the information that the data contains on the function $u_{\us}$ that the system optimizes is quantified by $\hat H[K]$. This conclusion is consistent with the fact that $\hat H[K]/\log 2$ is the (minimal) number of bits per state that is necessary to optimally encode the output of the experiment (see Ref.~\cite{Cov} Chap. 5). 

In order to gain intuition,  it is instructive to consider the case of extreme under sampling where each state is samples at most once, i.e. 
$K_{\us}=1$ for all states $\us$ in the sample and $K_{\us}=0$ otherwise. This corresponds to consider the regime $\beta\approx 0$ in Eq.\ (\ref{Logit}), where the data does not allow us to distinguish different observations in the sample and yields a uniform distribution on $\us$. 
At the other extreme, when the same state $\us_0$ is observed $M$ times, i.e. $K_{\us}=M\delta_{\us,\us_0}$, the data samples the function $u_{\us}$ in just one point $\us_0$.
In both cases the statistical range of the observed $K_{\us}$ does not allow us to learn much on the function $u_{\us}$ that is optimized.  Notice that $\hat H[K]=0$ in both these extreme cases, whereas 
$\hat H[\us]=\log M$ in the first case and $\hat H[\us]=0$ in the latter. 
Our intuition that in both these extreme cases we do not learn anything on the behavior of the system is 
precisely quantified by the value of $\hat H[K]$\footnote{To get an intuitive understanding of the information content of the two variables, imagine you want to find Mr $X$ in a population of $M$ individuals (this argument parallels the one in Ki Baek {\em et al.}~\cite{Baek}). Without any knowledge, this requires $\log M$ bits of information. But if you know that Mr $X$ lives in a city of size $k$, then your task is that of finding one out of $k \cdot m_k$ individuals, which requires $\log (k m_k)$ bits. Averaging over the distribution of $K$, we find that the information gain is given by $\hat H[K]$. 
How informative is the size of the city? Clearly if all individuals live in the same city, e.g. $m_k=\delta_{k,M}$, then this information is not very useful. At the other extreme, if all cities are formed by a single individual, i.e. $m_k=M\delta_{k,1}$, then knowing the size of the city where Mr $X$ lives is of no use either. In both cases $\log[k m_k]=\log M$. Therefore there are distributions $m_k$ of city sizes that are more informative than others. Notice that, in any case, the size $k$ of the city cannot provide more information than knowing the city $\us$ itself, i.e. $\hat H[K]\le \hat H[\us]$.}. For intermediate cases,
$\hat H[\us]$ will take an intermediate value in $[0,\log M]$ and we expect that different distributions are possible, which might provide a positive amount of information $\hat H[K]> 0$ on the system's behavior. 
Notice that, $K_{\us}>K_{\us'}$ suggests that state $\us$ is optimal under broader conditions than $\us'$. But if $K_{\us}=K_{\us'}$ the sample does not allow to distinguish the two states. In this sense, $\hat H[K]$ quantifies the number of states that the sample allows us to distinguish.

\subsection{Most informative samples}

Observation (2) above states that the distribution $p_{\us}$ can be seen as a distribution of maximal $E_{\us}[u]=\sum_{\us}p_{\us}u_{\us}$ with a given $H[\us]=\bar H$. 
The choice of which and how many variables to model, effectively fixes the number of unknown variables, that controls the inverse temperature parameter $\beta$ in Eq.~(\ref{Logit}), and ultimately tunes the entropy $\bar H$ to different values between zero and $n \log 2$.
Since\footnote{This follows from the Asymptotic Equipartition Property (AEP)~\cite{Cov} that derives from the law of large numbers and states that, when $M\gg 1$ is large
\[
-\frac{1}{M}\log P\{\us^{(1)},\ldots,\us^{(M)}\}=-\frac{1}{M}\sum_{i=1}^M \log p_{\us^{(i)}}\simeq H[\us].
\]
Using $P\{\us^{(1)},\ldots,\us^{(M)}\}=p_{\us^{(1)}}\cdots p_{\us^{(M)}}$, this leads to
\[
\hat H[\us]+D_{KL}(\hat p||p)\simeq H[\us],
\]
where $\hat p_{\us}=K_{\us}/M$ and $D_{KL}(\hat p||p)=\sum_{\us}\hat p_{\us}\log (p_{\us}/\hat p_{\us})$ is the Kullback-Leibler divergence.
Note that $\hat H[\us]\le \log M$, so if $M$ is not large enough $\hat H[\us]$ is not a good estimate of $H[\us]$. Since $D_{KL}(\hat p||p)\ge 0$, then $\hat H[\us]\le H[\us]$.} $\hat H[\us]\le H[\us]$, 
 we should look at empirical distributions with  bounded $\hat H[\us]\le \bar H$. Among these, those with maximal information content are those whose distribution $\mathbf{m}=\{m_k,~k>0\}$ is such that $\hat H[K]$ is maximal\footnote{A similar argument can be found in Baek  {\em et al.}~\cite{Baek}, though the analysis and conclusions presented here  differ substantially from those Ref.~\cite{Baek}.}:
\begin{equation}
\label{ }
\mathbf{m}^*={\rm arg}\max_{\mathbf{m}: \hat H[\us]\le \bar H}\hat H[K]
\end{equation}
subject to the additional constraint $\sum_k k m_k=M$. The solution to this problem is made non-trivial by the fact that $m_k$ should be a positive integer. Here we explore the solution within a very rough approximation where we consider $m_k$ a positive real number. This provides an upper bound to the entropy $\hat H[K]$ that we combine with the upper bound $\hat H[K]\le \hat H[\us]$ implied by the data processing inequality~\cite{Cov}, that arises from the fact that the random variable $K_{\us}$ is a function of $\us$.

\begin{figure} 
\begin{center} 
\includegraphics[width=6cm]{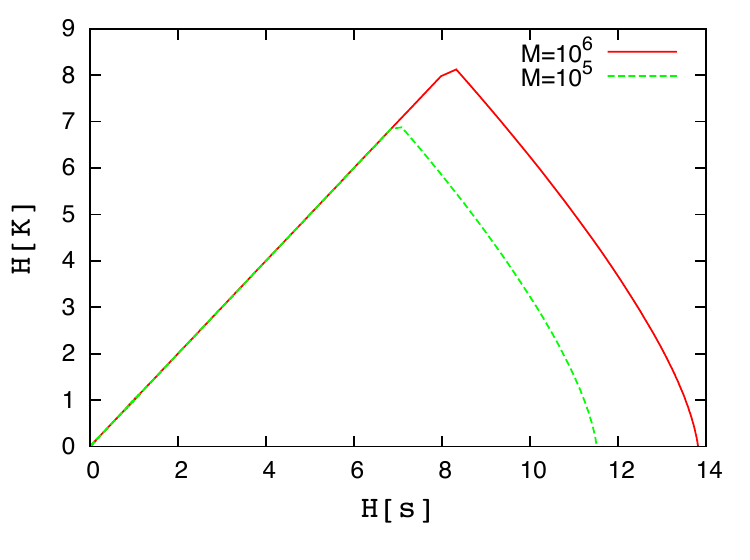} 
\includegraphics[width=6cm]{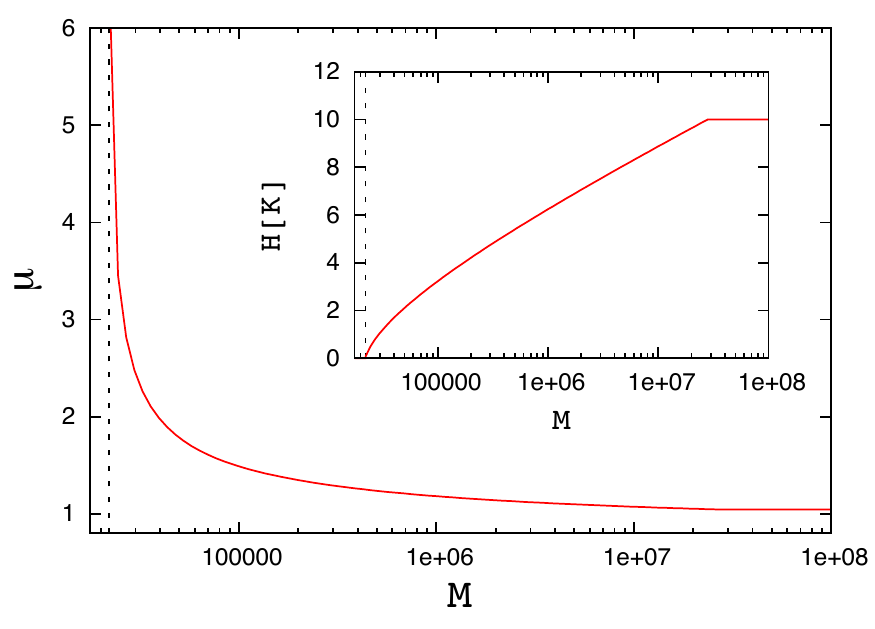} 
\caption{\label{HKHs} (left) Maximal entropy $\hat H[K]$ plotted as a function of the system entropy $\hat H[\us]$ for $M=10^5$ and $10^6$. The under sampled regime corresponds to the right region, while the left region for which $\hat H[K]\approx \hat H[s]$ represents the regime in which the distribution $p_\us$ is well sampled. The peak separating the two regimes is associated with a Zipf distribution for $m_k$. (right) Exponent $\mu$ as a function of $M$ within the approximated solution presented in the text, with $H[\us]=10$. In the inset we represent $\hat H[K]$ as a function of $M$. The vertical dashed line corresponds to $\log M=H[\us]$.} 
\end{center} 
\end{figure} 

In the region where $\hat H[K]<\hat H[\us]$, the solution to the approximated problem is readily found maximizing
\begin{equation}
\label{ }
\hat H[K]+\mu \hat H[\us]+\lambda\sum_{k>1} k m_k
\end{equation}
over $m_k\in \mathbb{R}^+$, where $\mu$ and $\lambda$ are Lagrange multipliers that are used to enforce the constraints $\hat H[\us]=\bar H$ and $\sum_{k=1}^M km_k=M$. The solution reads:
\begin{equation}
\label{expNk}
m_k^* = c k^{-1-\mu},\qquad 1\le k\le M
\end{equation}
where $c>0$ is a constant that is adjusted in order to enforce normalization. As $\mu$ varies, the upper bound draws a curve in the $\hat H[K]$ vs $\hat H[\us]$ plane, as shown in Fig.~\ref{HKHs} (left) for two values of $M$. In particular, the slope of the curve is exactly given by $-\mu$. Therefore we see that at the extreme right, $\hat H[K]\to 0$ as $\hat H[\us]\to\log M$ with infinite slope $\mu\to\infty$, corresponding to a distribution $m_k=M\delta_{k,1}$. As $\mu$ decreases, the distribution $m_k$ spreads out and  $\hat H[K]$ increases accordingly.

There is a special point where the upper bound $\hat H[K]$ derived from the solution with $m_k\in \mathbb{R}$ matches the data processing inequality line $\hat H[\us]=\hat H[K]$. We find that the slope of the line at this point (see Fig.~\ref{HKHs}) approaches $\mu=1$ from above, which corresponds to a distribution $m_k\sim k^{-2}$. 

In the regime where $\hat H[K]< \hat H[\us]$, the true distribution $p_{\us}$ is under sampled and a number of states $\us$ are all sampled an equal number of times. When $\hat H[K]= \hat H[\us]$, instead,  almost all states are sampled a different number of times. Therefore knowing the frequency $K_{\us}/M$ of a state is equivalent to knowing the state $\us$ itself. Notice that in this regime, $m_k$ is {\em not} given by the solution of the above optimization problem, since $\hat H[K]$ is bound by the data processing inequality. Indeed, in this regime, the empirical distribution converges to whatever the  underlying distribution is\footnote{There is an interesting duality between the distribution of $\us$ and that of $K$: When the former is under sampled (e.g. all states are seen only few times) the distribution $m_k$ is well sampled (i.e. $m_k\propto M$), whereas when ${\us}$ is well sampled, $m_k$ is under sampled, i.e. $m_k=0$ or $1$.}, with $m_k=0$ or $1$ for almost all the values of $k$. 

These results provide a picture of how most informative samples behave as the sample size $M$ increases, and the curve in the left part of Fig.~\ref{HKHs} moves upward (see Fig.~\ref{HKHs} right). 
As long as $\log M$ is smaller than the entropy $\bar H$ of the unknown distribution, we expect that all states in the sample will occur at most once, i.e. $\hat H[K]=0$. When $M\approx e^{\bar H}$, we start sampling states more than once. Beyond this point, $\hat H[K]$ will increase and $m_k\sim k^{-1-\mu}$ will take a power law form, with an exponent that decreases with $M$ (see Fig.~\ref{HKHs} right). When $M$ is large enough the entropy $\hat H[K]$ will saturate to the value $\bar H$ of the underlying distribution and $\mu$ will draw closer to one. Further sampling will provide closer and closer approximation of the true distribution $p_{\us}$  (see Fig.~\ref{subsamp_city}). 

The above argument suggests that power law distributions are the frequency distributions with the largest information content \emph{in the under sampled regime} (i.e. to the right of the cusp in Fig.~\ref{HKHs} left). The value of the exponent $\mu$ can be read from the slope of the curve. The maximum, that corresponds to a cusp, has $\mu\simeq 1$, hence a distribution that is close to the celebrated Zipf's law $m_k\sim k^{-2}$. Actually, the plot of $\mu$ vs $M$ in Fig.~\ref{HKHs} suggests that there is a broad range of $M$ over which $\mu$ takes values very close to one.


\subsection{Criticality and Zipf's law}

The results above suggest that Zipf's law ($\mu=1$) emerges as the most informative distribution which is compatible with a fixed value of the entropy $H[\us]$. Here we want to show how this is consistent with the approach in Ref.~\cite{MoraBialek}.
Mora and Bialek~\cite{MoraBialek} draw a precise relation between the occurrence of Zipf's law and criticality in statistical mechanics. In brief, given a sample and an empirical distribution $\hat p_{\us}=K_{\us}/M$ it is always possible to define an energy function $E_{\us}=-\log \hat p_{\us}$ and a corresponding entropy, $\Sigma(E)$ through the usual relation $e^{\Sigma(E)}=\frac{d\mathcal{N}(E)}{dE}$ with the number $d\mathcal{N}(E)$ of energy states between energy $E$ and $E+dE$. For $E=-\log (k/M)$, $d\mathcal{N}(E)=m_k\left|\frac{dk}{dE}\right|=k m_k$. Therefore, $\Sigma(E)=\log (k m_k)$ which means that Zipf's law $m_k\sim k^{-2}$ corresponds to linear relation $\Sigma(E)\simeq \Sigma_0+\beta E$ with slope $\beta=1$. The relation with criticality in statistical mechanics arises because the vanishing curvature in $\Sigma(E)$ corresponds to an infinite specific heat~\cite{MoraBialek}. 

The linearity of the $\Sigma(E)$ relation is not surprising. Indeed, the range of variation of entropy and energy in a sample of $M$ points is limited by $\delta \Sigma,\delta E\le \log M$. For intensive quantities $\sigma=\Sigma/n$ and $\epsilon=E/n$, this corresponds to a linear approximation of the $\sigma(\epsilon)\simeq \sigma_0+ \beta \epsilon$ relation over an interval $\delta \sigma,\delta \epsilon\sim (\log M)/n$ that can be relatively small. 
The fact that the coefficient takes the particular value $\beta\approx 1$ is, instead, non-trivial and it corresponds to the situation where the entropy vs energy relation enjoys a wider range of variation.

The results of the previous section\footnote{We remark an interesting formal analogy between the picture above and the statistical mechanics analogy of Ref.~\cite{MoraBialek}, within the simplified picture provided by our approximation. Upon defining $Z_\mu=\sum_k k^{-\mu}$, it is easy to check that $\hat H[\us]=\log M+\partial_\mu\log Z_\mu$ and $\hat H[K]=\log Z_\mu-\mu \partial_\mu\log Z_\mu$. Thus, identifying $Z_\mu$ with a partition function, $\hat H[\us]$ and $\hat H[K]$ stand precisely in the same relation as the energy and the entropy of a statistical mechanical system.}
provide an alternative perspective on the origin of Zipf's law: imagine a situation where we can choose the variables $\us$ with which to probe the system. Each choice corresponds to a different function $u_{\us}$ or to a different $\sigma(\epsilon)$ relation, of which the sample probes a small neighborhood of size $(\log M)/n$. For each choice of $\us$, this relation will likely look linear $\sigma(\epsilon)\simeq \sigma_0+\beta \epsilon$ with a different coefficient $\beta$. How should one choose the variables $\us$? It is clear that probing the system along variables for which $\beta\ll 1$ results in a very noisy dataset whereas if $\beta\gg 1$ one would be measuring constants. On the contrary, probing the system 
on ``critical'' variables, i.e. those for which $\beta\approx 1$, provides more information on the system's behavior. Zipf's law, in this perspective, is a consequence of choosing the known variables as those that reveal a wider range of variability in the $\sigma(\epsilon)$ relation.

\section{Applications}

Are the findings above of any use? 

\begin{figure} 
\begin{center} 
\includegraphics[width=9cm]{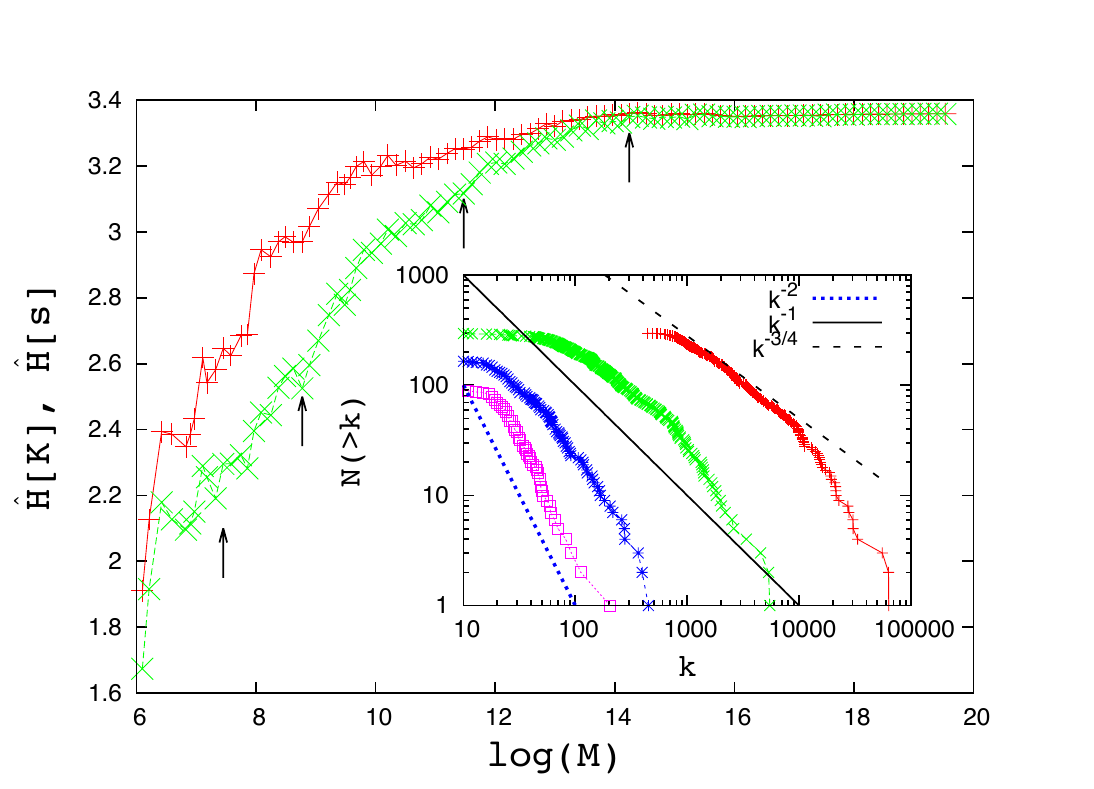} 
\caption{\label{subsamp_city} Distribution in cities for subsamples of $M$ households
of the IPUM database ({\tt http://usa.ipums.org}). Main figure: $\hat H[\us]$ and $\hat H[K]$ as function of $M$. Inset: cumulative distribution $N(>k)=\sum_{q>k}m_q$ of city distribution for subsamples of $M=1721, 6452, 96118$ and  $1535956$ (from left to right, corresponding to the  arrows in the main figure.} 
\end{center} 
\end{figure}

As we have seen, the distribution $m_k$ conveys information on the internal self-organization of the system. In the case of city size distribution, the occurrence of a broad distribution suggests that the city $\us$ is a relevant variable that enters in the  optimization problem that individuals solve. Indeed, individuals could be clustered according to different criteria (electoral districts, population living in areas of equal size, etc) and we don't expect broad distributions in general. 
Furthermore, we expect that if we progressively sample a population of individuals, the resulting city size distribution would ``evolve'' approximately as described above. Fig.~\ref{subsamp_city} shows the result of such an exercise for a data set of US citizens (see caption). 
Interestingly, we find that for small samples the distribution takes a power law form $m_k\sim k^{-\mu-1}$ with exponent $\mu>1$, and as $M$ increases the distribution gets broader (i.e. $\mu$ decreases) and converges to the city size distribution, when only 0.5\% of the individuals are sampled\footnote{Cristelli {\em et al.}~\cite{pnerozipf} have shown that Zipf's law does not hold if one restricts the statistics to a subset of cities which is different from the set over which self-organization takes place. This points to a notion of {\em coherence} of the sample, which is consistent with our framework where the sample is thought of being the outcome of an optimization problem. Note that our subsampling differs from the one in Ref.~\cite{pnerozipf} as we are sampling individuals rather than cities.}.

In most applications the relevant variables are not known. In this case, the maximization of $H[K]$ can be used as a guiding principle to select the most appropriate variables or to extract them from the data. We illustrate the problem with three examples.

%
%
%

\subsection{Protein sequences}

A protein is defined in terms of its amino-acid sequence\footnote{Each $s_i$ takes $21$ values rather than $2$, but that is clearly an non-consequential difference with respect to the case where $s_i=\pm 1$.} $\vec s$ but its functional role in the cell, as well as its 3d structure, is not easily related to it. The sequences $\vec s$ of homologous proteins -- i.e. those that perform the same function -- can be retrieved from public databases~\cite{Sanger}. Mutations across sequences of homologous proteins are such that they preserve that function but otherwise might be optimized in order to cope with their particular cellular environment. 
 This suggests that there may be relevant amino-acids $\underline{s}$, that are optimized for preserving the function and less relevant ones. 

\begin{figure} 
\begin{center} 
\includegraphics[width=9cm]{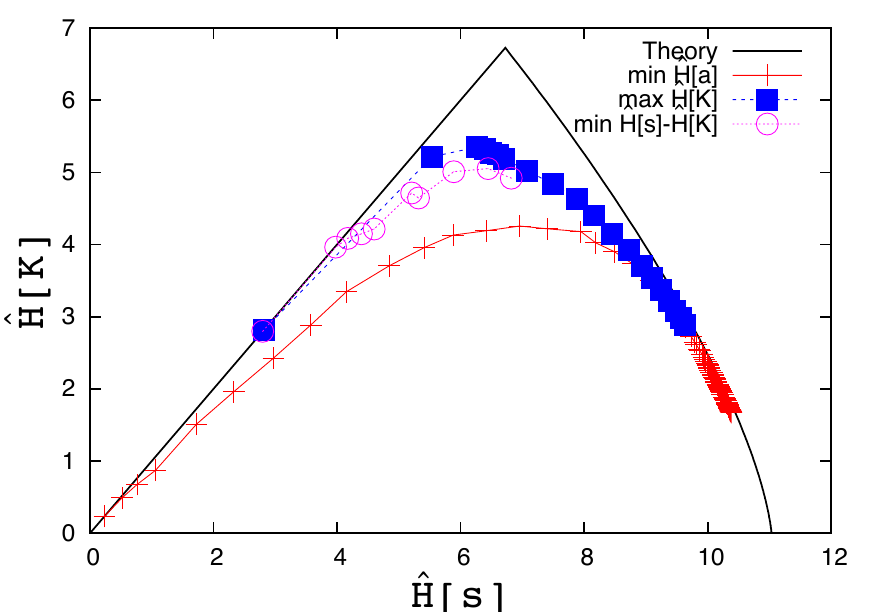} 
\caption{\label{HkHsProt}Entropy  $\hat H[K]$ as a function of $\hat H[\underline{s}]$ for the protein family PF000072. Subsequence of the $n$ most conserved positions (red $+$); Subsequences of $n$ positions with maximal $\hat H[K]$ (blue $\blacksquare$) and with minimal $\hat H[\us]-\hat H[K]$ (pink $\circ$). 
$n$ increases from left to right in all cases.} 
\end{center} 
\end{figure} 

How to find relevant variables? One natural idea is to look at the subsequence of the $n$ most conserved amino acids\footnote{For any given subset $\underline{s}$ of the $\vec{s}$ variables, the frequency $\hat p_{\underline{s}}$ can computed  and, from this the entropies $\hat H[\us]$ and $\hat H[K]$. As a measure of conservation, we take the entropy  of the empirical distribution of amino acids in position $i$.}. Fig.~\ref{HkHsProt} shows the information content $\hat H[K]$ as a function of $\hat H[\us]$ as the number $n$ of ``relevant'' amino acids varies for the family PF000072 of response regulator receiver proteins\footnote{Our analysis is based on $M=62074$ sequences, that after alignment, are $N=112$ amino-acids long. The same data was used in Ref.~\cite{Wegt}.}~\cite{Sanger}.
For $n$ large, most of the sequences are seen only once (small $\hat H[K]$), and $\hat H[\underline{s}]\propto \log M$, whereas for $n<25$ the entropy $\hat H[\underline{s}]$ decreases steeply as $n$ decreases. Correspondingly, $\hat H[K]$ exhibits a maximum at $n=n_c=22$ and then approaches $\hat H[\us]$.

\begin{figure} 
\begin{center} 
\includegraphics[width=4.5cm]{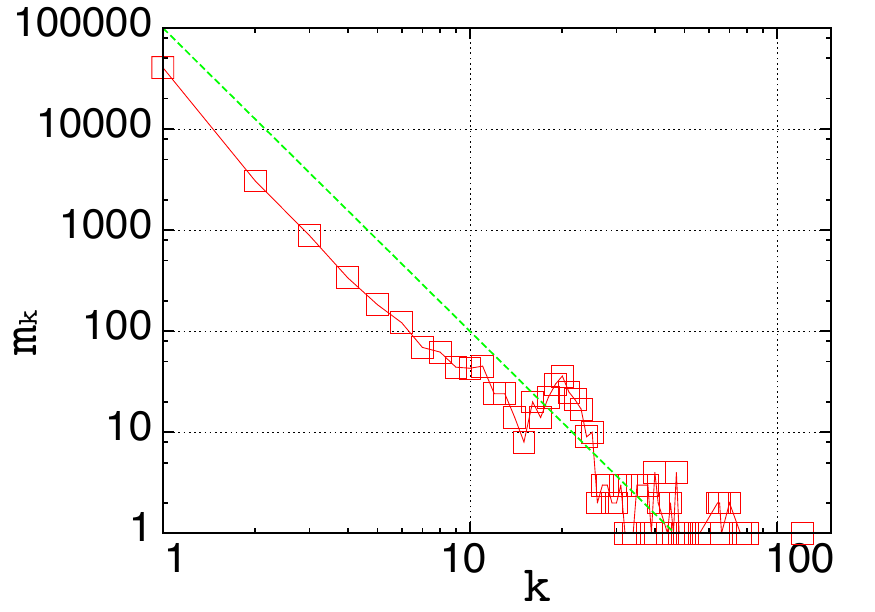} 
\includegraphics[width=4.5cm]{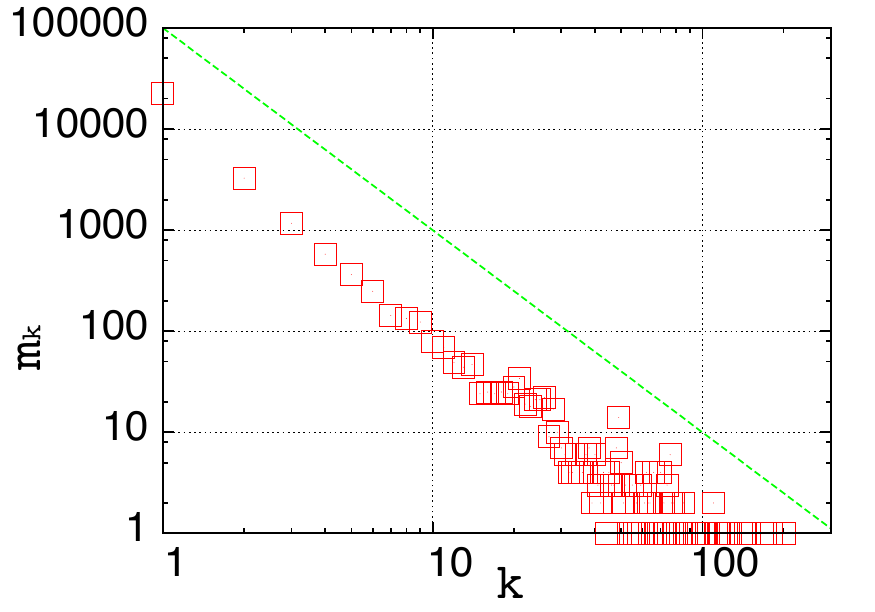} 
\includegraphics[width=4.5cm]{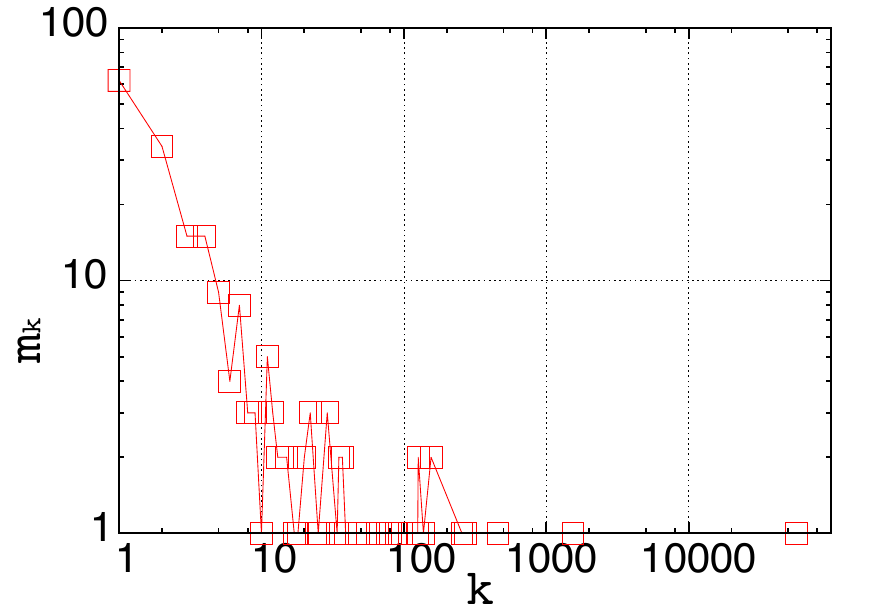} 
\caption{\label{figzipf} Frequency distribution $m_k$ for $n=N=112$ (left), $n=22\approx n_c$ (center) and $n=2$ (right). Lines are proportional to $k^{-3}$ (left) and $k^{-2}$ (center).} 
\end{center} 
\end{figure}

Even if the empirical curve does not saturate the theoretical bound, the frequency distribution exhibits Zipf's law around the point $n_c$ where $\hat H[K]$ is maximal. Fig.~\ref{figzipf} shows that for $n\approx n_c$ the number $m_k$ of sequences that are sampled $k$ times falls off as $m_k\sim k^{-2}$, characteristic of a Zipf's law, whereas for $n\approx N$ it falls off faster and for $n\sim O(1)$ it is dominated by one large value of $k\approx M$. 

Alternatively, one may use the maximization of $\hat H[K]$ as a guide for identifying the relevant variables. We do this by an agglomerative algorithm, where we start from a sequence $\us$ of length zero and iteratively build subsequences of an increasing number $n$ of sites. At each step, we add the site $i$ that makes the information content $\hat H[K]$ of the resulting subsequence as large as possible\footnote{Notice that the algorithm is not guaranteed to return the subset of sites that maximizes $\hat H[K]$ for a given $n>1$.}. The result, displayed in Fig.~\ref{HkHsProt}, shows that this procedure yields subsequences with an higher $\hat H[K]$ which are also shorter. In particular, the maximal $\hat H[K]$ is achieved for subsequences of just three amino acids. 

Interestingly, if one looks at the subsequence of sites that are identified by this algorithm one finds that the first two sites of the subsequence are among the least conserved ones:  they are those that allow to explain the variability in the dataset in the most compact manner -- loosely speaking, they are ``high temperature''  variables ($\beta\ll 1$). The following ten sites identified by the algorithm are instead ``low temperature" variables, as they are the most conserved ones. This hints at the fact that relevant variables should not only encode a notion of optimality, but also account for the variability within the data set, under which the system is (presumably) optimizing its behavior. 

\subsection{Clustering and correlations of financial returns}

In many problems data is  noisy and high dimensional. It may consist of $M$ observations $\hat x=(\vec x^{(1)},\ldots,\vec x^{(M)})$ of a vector of features $\vec x\in \mathbb{R}^T$ of the system under study. Components of $\vec x$ may be continuous variables, so the analysis of previous sections is not applicable. In these cases a compressed representation $\us^{(i)}$ of each point $\vec x^{(i)}$ would be desirable, where $\us$ takes a finite number of values and can be thought of as encoding a relevant description of the system. There are several ways to derive a mapping $\us=F(\vec x)$, such as quantization~\cite{Cov} or data clustering. The general idea is that of discretizing the space of $\vec x$ in cells, each labeled by a different value of $\us$, so ``similar'' points $\vec x^{(i)}\approx \vec x^{(j)}$ fall in the same cell, i.e. $\us^{(i)}=\us^{(j)}$. The whole art of data clustering resides in what ``similar'' exactly means, i.e. on the choice of a metrics in the space of $\vec x$. 
Different data clustering algorithms differ on the choice of the metrics as well as on the choice of the algorithm which is used to group similar objects in the same cluster and on the resolution, i.e. on the number of clusters. Correspondingly, different clustering algorithms extract a different amount of information on the internal structure of the system. 
In practice, how well the resulting cluster structure reflects the internal organization of the data depends on the specific problem, but there is no unambiguous manner, to the best of our knowledge, to compare different methods.

The point we want to make here is that the discussion of the previous section allows us to suggest an universal method to compare different data clustering algorithms and to identify the one that extracts the most informative classification. The idea is simple: For any algorithm A, compute the variables $K_{\us}^A$ and the corresponding entropies $\hat H[\us^A]$ and $\hat H[K^A]$ and plot  the latter with respect to the former, as the number $n$ of clusters varies from $1$ to $M$. If such curve for algorithm A lies above the corresponding curve for algorithm B, we conclude that A extracts more information on the systems behavior and hence it is to be preferred to B.

\begin{figure} 
\begin{center} 
\includegraphics[width=9cm]{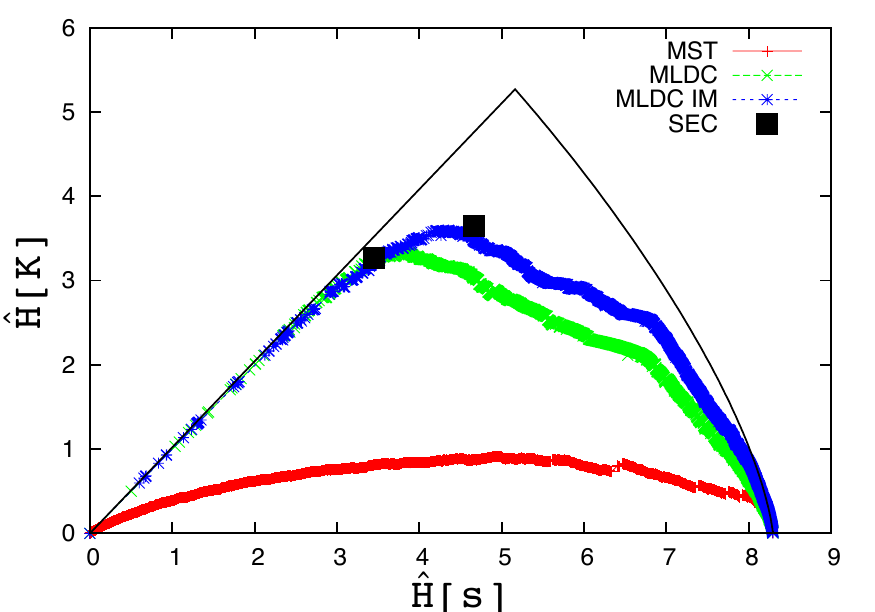} 
\caption{\label{HsHkFin}Entropy  $\hat H[K]$ as a function of $\hat H[\underline{s}]$ as the number $n$ of clusters increases (from left to right), for different data clustering schemes. From bottom to top, Single Linkage (MST), maximum likelihood with (MLDC) and without (MLDC IM) the principal component. The SEC classification at 2 and 3 digits of the stocks is also shown as black squares.} 
\end{center} 
\end{figure} 

This idea is illustrated by the study of financial correlations of a set of $M=4000$ stocks in NYSE in what follows\footnote{Here $\vec x^{(i)}=(x_1^{(i)},\ldots,x_T^{(i)})$ consists of daily log returns $x_t^{(i)}=\log (p_t^{(i)}/p_{t-1}^{(i)})$, where $p_t^{(i)}$ is the price of stock $i$ on day $t$, and $t$ runs from 1st January 1990 to 30th of April 1999.}.
Financial markets perform many functions, such as channelling private investment to the economy, allowing inter-temporal wealth transfer and risk management. Time series of the price dynamics carry a signature about such complex interactions, and have been studied intensively~\cite{rev_corr1,rev_corr2, micciche}: the principal component in the singular value decomposition  largely reflects portfolio optimization strategies whereas the rest of the correlations exhibit a structure which is highly correlated with the structure of economic sectors, down to a scale of 5 minutes~\cite{micciche}. 
Since we're borrowing this example to make a generic point, we shall not enter  into further details, and refer the interested reader to~\cite{rev_corr1,rev_corr2, micciche}. 
Several authors have applied Single Linkage data clustering method to this problem~\cite{rev_corr1}, which consists in building Minimal Spanning Trees where the links between the most correlated stocks, that do not close loops, are iteratively added to a forest. Clusters are identified by the disconnected trees that, as links are sequentially added, merge one with the other until a single cluster remains. The resulting curve $\hat H[K]$ vs $\hat H[\us]$ is shown in Fig.~\ref{HsHkFin}.

A different data clustering scheme has been proposed in Ref.~\cite{MLDC,micciche} based on a parametric model of correlated random walks for stock prices. The method is based on maximizing the likelihood with an hierarchical agglomerative scheme~\cite{MLDC}. The curve $\hat H[K]$ vs $\hat H[\us]$ lies clearly above the one for the MST (see Fig.~\ref{HsHkFin}). Ref.~\cite{micciche} has shown that the structure of correlation is revealed more clearly if the principal component dynamics is subtracted from the data\footnote{If $x^0_t$ is the principal component in the singular value decomposition of the data set, this amount to repeating the analysis for the modified dataset $\tilde x^{(i)}_t=x^{(i)}_t-x^0_t$.}. This is reflected by the fact that the resulting curve $\hat H[K]$ vs $\hat H[\us]$ shifts further upward. In the present case, it is possible to compare these results with the classification given by the U.S. Security and Exchange Commission (SEC), which is given by the black squares in Fig.~\ref{HsHkFin} for 2 and 3 digits SEC codes. This classification codifies the information on the basis of which agents trade, so it enters into the dynamics of the market. 
The curve obtained removing the principal component draws remarkably close to these points, suggesting that the clustering method extracts a large fraction of the information on the internal organization of the market. Again, the rank plot of cluster sizes reveals that Zipf's law occurs where $\hat H[K]$ is close to its maximum, whereas marked deviations are observed as one moves away from it.

\subsection{Keywords in a text}

A written text can be thought of as the result of a design, by the the writer: There are tens of thousands of words in the vocabulary of a given language, but in practice the choice is highly constrained by syntax and semantics,  as revealed by the fact that the frequency distribution in a typical text is highly peaked on relatively few words, and it roughly follows Zipf's law. 

The frequency with which a given ford $w$ occurs in a given section $\us$ of a manuscript should contain traces of the underlying optimization problem. This insight has been exploited by 
Montemurro and Zanette~\cite{MMDZ} in order to extract keywords from a text.
The idea in Ref.~\cite{MMDZ} is: {\em i)} split the text into parts $\us$ of $L$  consecutive words; {\em ii)} compute the fraction $\hat p_{\us}^{(w)}$ of times word $w$ appears in part $\us$; {\em iii)} compute the difference $\Delta H[\us]$ between the entropy $\hat H[\us]$ of a random reshuffling of the words in the parts and the actual word frequency. 
Keywords are identified with the least random words, those with the largest $\Delta H[\us]$.

\begin{figure} 
\begin{center} 
\includegraphics[width=9cm]{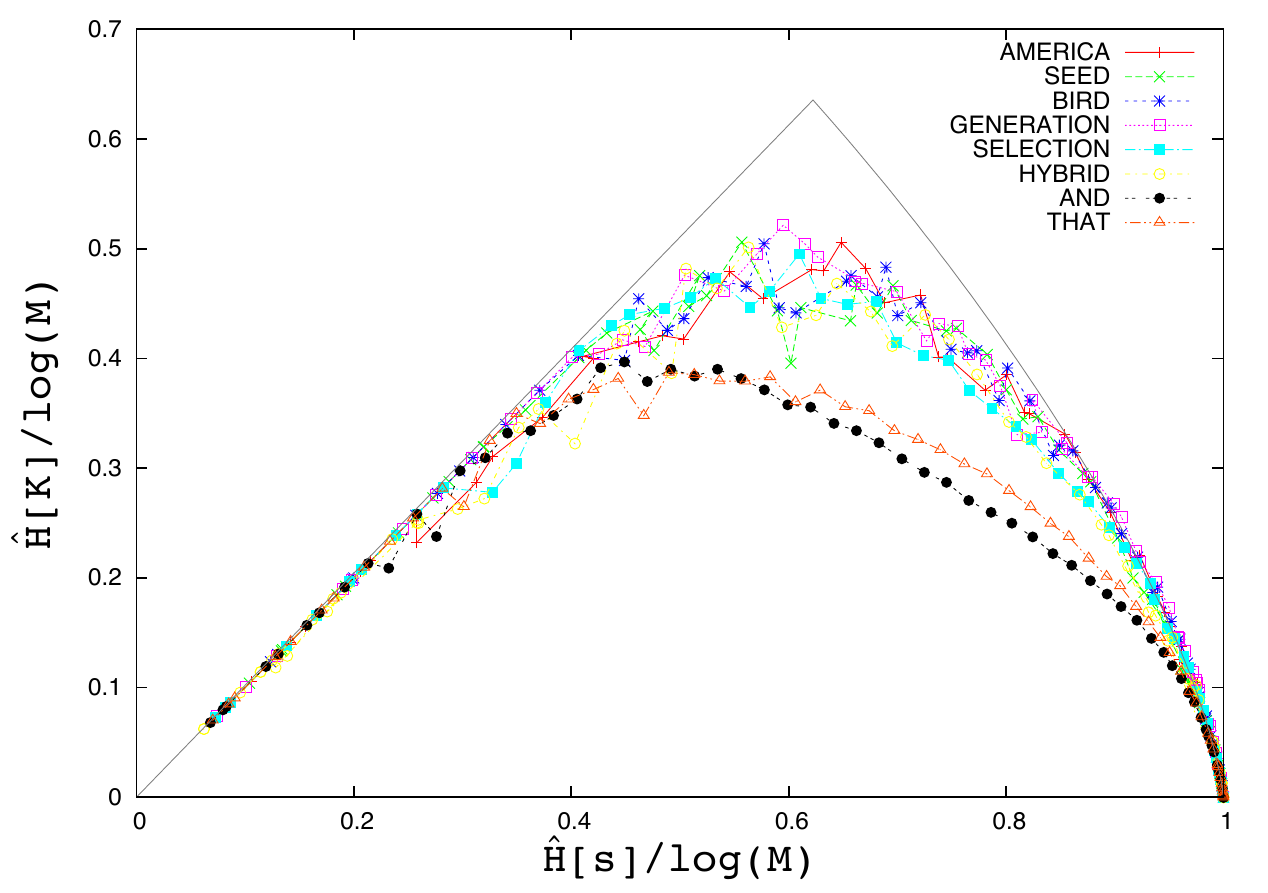} 
\caption{\label{hkhsw}Entropy  $\hat H[K]$ as a function of $\hat H[\underline{s}]$ for the occurrence of different words (see legend) of Darwin's  "On the Origin of Species'' in segments of $L$ consecutive words ($L$ increasing from right to left).} 
\end{center} 
\end{figure}

From our perspective, for each choice of $L$ and each word $w$, one can compute $\hat H^w[K]$ and $\hat H^w[\us]$. Fig.~\ref{hkhsw} shows the resulting curve as $L$ varies for Darwin's ``On the Origin of Species''. Among all words that occur at least 100 times, we select those that achieve a maximal value of $\hat H[K]$ as well as some of those whose maximal value of $\hat H[K]$ (on $L$) is the smallest. The latter turn out to be generic words (``and'', ``that'') whereas among the former we find words (e.g. ``generation'', ``seed", ``bird'') that are very specific of the subject discussed in the book. Whether this observation can be used to derive a more efficient extractor of keywords than the one suggested in Ref.~\cite{MMDZ} or not, is a question that we leave for future investigations. For our present purposes, we merely observe that $\hat H[K]$ allows us to distinguish words that are ``mechanically'' chosen from those that occur as a result of a more complex optimization problem (the keywords).

\section{Discussion}

Advances in IT and experimental techniques have boosted our ability to probe complex systems to unprecedented level of detail. Increased performance in computing, at the same time, has paved the way to reproducing  {\em in silico} the behavior of complex systems, such as cells~\cite{simcell},  the brain~\cite{connectome} or the economy~\cite{eurace}. 

However it is not clear whether this approach will ultimately deliver predictive models of complex systems. Interestingly, 
Ref.~\cite{machine_learning} observes that efforts in Artificial Intelligence to reproduce {\em ab initio} human capabilities in intelligent tasks have completely failed: Search engines, recommendation systems and automatic translation~\cite{machine_learning} have been achieved by unsupervised statistical learning approaches that harvest massive data sets, abandoning altogether the ambition to understand the system or to model it in detail.
At the same time, problems such as drug design~\cite{drugdesign} and the regulation of financial markets~\cite{Haldane} still remain elusive, in spite of increased sophistication of techniques deployed.

This calls for understanding the limits of modeling complex systems and devising ways to select relevant variables and compact representations. The present contribution is an attempt to address these concerns. In doing that, we uncover a non-trivial relation between ``criticality'', which in this context is used to refer to the occurrence of broad distributions in the frequency of observations (Zipf's law), and the relevance of the measured variables. We make this relation precise by quantifying the information content of a sample: Most informative data, that sample relevant variables, exhibit power law frequency distributions, in the under sampling regime. 
Conversely, a description in terms of variables which are not the ones the system cares about will not convey much information. Mostly informative data set are those for which the frequency of observations covers the largest possible dynamic range, providing information on the system's optimal behavior in the wider range of possible circumstances. This corresponds to a linear entropy-energy relation, in the statistical mechanics analogy discussed in Ref.~\cite{MoraBialek}.


Our results point in the same direction of the recent finding that inference of high dimensional models is likely to return models that are poised close to ``critical'' points~\cite{IMMM}. This builds on the observation~\cite{VB} that the mapping between the parameter space of a model and the space of distributions can be highly non-linear. In particular, it has been shown in simple models~\cite{IMMM} that regions of parameter space of models that have a vanishing measure (critical points) concentrate a finite fraction of the possible (distinguishable) empirical distributions. This suggests that ``optimally informative experiments'' that sample uniformly the space of empirical distributions are likely to return samples that look ``close to a critical point'' when we see them through the eyes of a given parametric model. 


Our findings are also consistent with the observation~\cite{pnerozipf} that Zipf's law entails some notion of ``coherence of the sample'' in the sense that typical subsamples deviate from it. In our setting, the characteristics that makes the sample homogeneous is that it refers to systems ``doing the same thing'' under ``different conditions''. 

As shown in the last section, the ideas in this paper can be turned into a criterium for selecting mostly informative representations of complex systems. This, we believe, is the most exciting direction for future research. One particular direction in which our approach could be useful is that of the identification of hidden variables, or {\em unknown unknowns}. 
In particular, the identification of relevant classification of the data can be turned into the specification of hidden variables, whose interaction with the observed ones can be inferred. 
This approach would not only predict how many hidden variables should one consider, but also how they specifically affect the system under study. 
Progress along these lines will be reported in future publications.

\section*{Acknowledgements}
We gratefully acknowledge William Bialek, Andrea De Martino, Silvio Franz, Thierry Mora, Igor Prunster, Miguel Virasoro and Damien Zanette for various inspiring discussions, that we have taken advantage of.

\appendix

\section{When are models predictive? The Gaussian case}

In this appendix, we consider the setup of Section \ref{sec:setup} in the case of a Gaussian distribution of $v_{\bar s|\us}$, for which $\beta=\sqrt{2N(1-f)\log 2}$. Here and in the rest of the appendix $f=n/N$ is the fraction of known variables and we shall focus on the the asymptotic behavior in the limit $n,N\to\infty$ with $f=n/N$ finite.

We assume that the dependence of the objective function $u_{\us}$ on known variables $\us=(s_1,\ldots,s_n)$ is known and we 
concentrate on the specific example where $u_{\us}$ are also i.i.d. draws from a Gaussian distribution with zero mean and variance $\sigma^2$.  
This is the most complex system one could think of, as its specification requires an exponential number of parameters. 
As argued in section \ref{sec:Gibbs}, this is also a particular case where the subset of known variables coincides with the subset of the most relevant ones.
The question we address is: does the knowledge of the function $u_{\us}$ allows us to predict the optimal behavior $\us^*$? 

As a prototype example, consider the problem of reverse engineering the choice behavior of an individual that is optimizing an utility function $U(\vec s)$. For a consumer, $\vec s$ can be thought of as a consumption profile, specifying whether the individual has bought good $i$ ($s_i=+1$) or not ($s_i=-1$) for $i=1,\ldots,N$. However, consumer behavior can be observed only over a subset ${\underline{s}}=(s_1,\ldots, s_n)$ of the variables, and only the part $u_{\underline{s}}$ of the utility function that depends solely on the observed variables can be modeled\footnote{This setup is the one typically considered in random utility models of choice theory in economics~\cite{McFadden}.}.
Under what conditions the predicted choice ${\underline{s}}_0$ is informative on the actual behavior ${\underline{s}}^*$ of the agent?
Put differently, how relevant and how many (or few) should the relevant variables be in order for $\underline{s}_0$ 
to be informative on the optimal choice $\underline{s}^*$?

In light of the result of section \ref{sec:Gibbs}, the answer depends on how peaked is the distribution $p_{\us}$. 
For $\beta\to\infty$ the probability distribution concentrates on the choice ${\underline{s}}_0$ that maximizes $u_{\underline{s}}$ whereas for $\beta\to 0$ it spreads uniformly over all $2^n$ possible choices ${\underline{s}}$. 
Our problem, in the present setup, reverts to the well known REM, that is discussed in detail e.g. in Refs.~\cite{REM,MM}. We recall here the main steps.

The entropy of the distribution $p_{\underline{s}}$ is given by:
\begin{equation}
\label{eqH}
H[{\underline{s}}]=-\sum_{\underline{s}}p_{\underline{s}}\log p_{\underline{s}}=\log Z(\beta)-\beta\frac{d}{d\beta} \log Z(\beta)
\end{equation}
where the last equality is easily derived by a direct calculation.

In order to estimate $Z(\beta)$ let us observe that $2^{-n} Z(\beta)$ is an average and the law of large numbers suggests that it should be close to the expected value of $e^{\beta u_{\underline{s}}}$
\begin{equation}
\label{llnZ}
\frac{1}{2^n}Z(\beta)
\simeq E\left[ e^{\beta u_{\underline{s}}}\right]=e^{\beta^2\sigma^2 /2}\equiv \frac{1}{2^n} Z_{\rm ann}(\beta) 
\end{equation}
that depends on the fact that $u_{\underline{s}}$ is a Gaussian variable with zero mean and variance $\sigma^2 $.
Therefore, if we use $Z_{\rm ann}$ instead of $Z$ in Eq.~(\ref{eqH}), we find
\begin{equation}
\label{ }
H[{\underline{s}}]\simeq  n\log 2-\frac{\beta^2 \sigma^2 }{2}=N\left[ f-(1-f)\sigma^2 \right] \log 2.
\end{equation}
One worrying aspect of this result is that if 
\begin{equation}
\label{sigmac}
\sigma\ge\sigma_c=\sqrt{\frac{f}{1-f}}
\end{equation}
the entropy is negative. 
The problem lies in the fact that the law of large number does not hold for $\sigma\ge\sigma_c$ due to the explicit dependence of $\beta$ on $N$, in the limit $N\to\infty$. In order to see this, notice that the expected value of $u_{\underline{s}}$ over $p_{\underline{s}}$ is given by
\begin{equation}
\label{avgus}
u_{\us^*}^{({\rm ann})} = \sum_{\underline{s}} p_{\underline{s}} u_{\underline{s}}=\frac{d}{d\beta}\log Z\simeq
\beta \sigma^2 
= \sigma^2\sqrt{2N(1-f)\log 2}
\end{equation}
where the second relation holds when the law of large numbers holds. However, this 
cannot be larger than the maximum of $u_{\underline{s}}$ which, again by extreme value theory of Gaussian variables, is given by
\begin{equation}
\label{maxus}
u_{\underline{s}_0}=\max_{\underline{s}} u_{\underline{s}}\simeq \sigma\sqrt{2Nf\log 2}.
\end{equation}
Indeed the estimate in Eq.~(\ref{avgus}) gets larger than the maximum given in Eq.~(\ref{maxus}) precisely when $\sigma\ge \sigma_c$, i.e. when $H[{\underline{s}}]$ becomes negative. It can be shown that the law of large numbers, and hence the approximation used above, holds only for $\sigma<\sigma_c$~\cite{REM,MM}. The basic intuition is that for $\sigma<\sigma_c$ the sum in $Z$ is dominated by exponentially many terms (indeed $e^{H[{\underline{s}}]}$ terms) whereas for $\sigma\ge \sigma_c$ the sum is dominated by the few terms with  $u_{\us} \simeq \max u_{\underline{s}}$.

For $\sigma<\sigma_c$ we can use Eq.~(\ref{llnZ}) and (\ref{maxus}) to compute
\begin{equation}
p_{\underline{s}_0}=P\{{\underline{s}_0}=  {\underline{s}^*}\}\simeq 
e^{-N(1-f)(\sigma-\sigma_c)^2},
\qquad \sigma^2<\sigma^2_c \, ,
\end{equation}
which is exponentially small in $N$. Therefore the model prediction $\underline{s}_0$ carries no information on the systems' behavior $\underline{s}^*$ for $\sigma<\sigma_c$.

On the other hand, for $\sigma>\sigma_c$, $Z(\beta)$ is dominated by $u_{\underline{s}_0}$ and it can be estimated expanding the number $\mathcal{N}(u)=2^n e^{-u^2/(2\sigma^2)}/\sqrt{2\pi \sigma^2}$ of choices ${\underline{s}}$ with  $u_{\us}=u$ around $u_{\underline{s}_0}$.
Simple algebra and asymptotic analysis reveals that
\begin{equation}
\label{ }
p_{\underline{s}_0}\simeq 1-\frac{\sigma_c}{2\sqrt{\pi f\log 2}(\sigma-\sigma_c)+\sigma_c}+O(N^{-1}).
\end{equation}
In words, the transition from the region $p_{\underline{s}_0}\simeq 0$ to the region where $p_{\underline{s}_0}\simeq 1$ is rather sharp, and it takes place in a region of order $|\sigma-\sigma_c|\sim 1/\sqrt{N}$.

The most remarkable aspect of this solution is that $\sigma_c$ increases with $f$: 
for a given value of $\sigma$ the correct solution $\underline u^*$ is recovered only if the fraction of known variables is {\em less} than a critical value 
\begin{equation}
\label{ }
f_c=\sigma^2/(1+\sigma^2)
\end{equation}
This feature is ultimately related to the fact that the effect of unknown unknowns is a decreasing function of the number $N(1-f)$ of them (see Eq.~(\ref{maxv})). This, in turn, is a consequence of the Gaussian nature of the variables $v_{\bar s|\underline{s}}$ or in general of the fact that the distribution of $u$ and $v$ falls off faster than exponential.

\end{document}